\documentclass[12pt]{article}
\usepackage[super,sort&compress]{natbib}
\usepackage[T1]{fontenc}
\usepackage[utf8]{inputenc}
\usepackage{lmodern}
\usepackage[margin=1in]{geometry}
\usepackage{graphicx}
\usepackage{amsmath,amssymb}
\usepackage{natbib}
\usepackage{url}
\usepackage{lineno}
\usepackage{setspace}
\usepackage{float}
\usepackage{caption}
\usepackage{hyperref}
\setcitestyle{numbers,square,sort&compress}
\newcommand{\includeoriginal}[2][]{%
\IfFileExists{#2}{\includegraphics[#1]{#2}}{%
\fbox{\parbox[c][1.3in][c]{0.86\linewidth}{\centering Original figure file not attached:\\ \texttt{\detokenize{#2}}\\ Upload to Figure/ to restore.}}}}
\title{Carbon dioxide responses shape estimates of Amazon forest vulnerability under solar geoengineering}
\author{Anupama K. Xavier$^{1,*}$ and Chris D. Jones$^{1,2}$\\[.5em]
\small $^1$School of Geographical Sciences, University of Bristol, UK\\
\small $^2$Met Office Hadley Centre, Exeter, UK\\
\small $^*$Correspondence: anupama.kuttikkatxavier@bristol.ac.uk}
\date{}
\begin{document}
\maketitle

\begin{abstract}
The Amazon rainforest is vulnerable to warming and changes in rainfall, yet whether solar geoengineering can reduce conditions associated with forest loss remains uncertain. Stratospheric aerosol injection (SAI) can offset greenhouse-gas-driven warming without fully restoring regional hydroclimate. Here we assess Amazon forest vulnerability under mitigation, high-emission and SAI scenarios using a CO$_2$-dependent forest-sustainability threshold in a common climate phase space. By late century, SAI reduces projected Amazon warming from 5.49~$^\circ$C under high emissions to 2.45~$^\circ$C, but substantial drying and longer dry seasons persist. Under SAI, 44.7\% of analysed grid cells exceed the present-day threshold, compared with 8.7\% when the future CO$_2$-dependent threshold is considered. This contrast reveals that apparent forest protection depends strongly on the assumed resilience benefits of elevated CO$_2$. Our findings highlight the importance of hydroclimatic changes and CO$_2$-dependent vegetation responses in assessing the potential of solar geoengineering to reduce Amazon forest vulnerability.
\end{abstract}

The Amazon rainforest is a critical component of the Earth system, supporting exceptional biodiversity, storing substantial amounts of carbon and regulating regional and global climate through land--atmosphere interactions \citep{betts2008future,levine2016ecosystem}. However, rising temperatures, reduced precipitation and longer dry seasons threaten forest resilience by increasing water stress and potentially promoting transitions from high-biomass tropical forest towards lower-biomass vegetation states \citep{malhi2009exploring,brando2025tipping}. These responses vary spatially, reflecting differences in climate, soil properties, vegetation characteristics and disturbances such as fire and deforestation \citep{levine2016ecosystem,brando2025tipping}.

Despite international efforts to limit global warming, existing mitigation pathways leave substantial risks of climate-driven ecosystem change \citep{ipcc2022wg3}. Moreover, exceeding temperature targets, even temporarily, may trigger long-lived or committed changes in vulnerable ecosystems. For the Amazon, forest responses to climate change may persist beyond the period of warming because of slow vegetation adjustment and land--atmosphere feedbacks \citep{jones2009committed,munday2025risks}.

Solar radiation modification (SRM) has been proposed as a potential means of reducing climate risks, although it cannot replace greenhouse gas mitigation. One widely studied approach, stratospheric aerosol injection (SAI), aims to offset warming by increasing the reflection of incoming solar radiation \citep{crutzen2006albedo,robock2009benefits}. However, reducing radiative forcing does not necessarily restore the climate associated with lower atmospheric CO$_2$ concentrations. In particular, temperature, precipitation and atmospheric circulation respond differently to greenhouse gas and aerosol forcing \citep{kravitz2013energetic,ipcc2021wg1}.

This distinction is especially important for tropical forests, whose persistence depends on both thermal and hydrological conditions. Geoengineering simulations indicate that SAI can substantially reduce greenhouse-gas-driven warming while producing heterogeneous precipitation responses, including changes in tropical rainfall and atmospheric circulation \citep{tilmes2013hydrological,simpson2019regional,wells2024identifying}. Consequently, reduced warming over the Amazon may not be accompanied by an equivalent reduction in dry-season stress. At the same time, elevated atmospheric CO$_2$ can modify vegetation responses through physiological processes, further complicating assessments of forest vulnerability \citep{good2011quantifying,levine2016ecosystem}.

Previous studies suggest that SAI could increase terrestrial carbon storage through reduced warming and elevated atmospheric CO$_2$ concentrations \citep{parry2026stratospheric}. However, increased carbon storage does not necessarily establish whether environmental conditions remain suitable for long-term forest persistence. The extent to which SAI reduces exposure to conditions associated with forest loss therefore remains uncertain, particularly when temperature, precipitation seasonality and CO$_2$-dependent vegetation responses are considered together.

Here, we assess Amazon climate states relative to a CO$_2$-dependent forest-sustainability threshold under mitigation, high-emission and SAI pathways. By expressing the conditions associated with equilibrium forest persistence in a common temperature--dry-season-length--CO$_2$ phase space, we evaluate projected changes across five Earth system models without directly transferring geographical patterns of vegetation change. We examine whether SAI-induced cooling reduces exposure to threshold-exceeding conditions and how this assessment depends on residual hydroclimatic changes and the CO$_2$-dependent vegetation response.

\section{Results}
\subsection{Equilibrium forest response and forest-sustainability threshold}

To assess how SAI influences climatic conditions associated with Amazon forest persistence, we combine equilibrium vegetation experiments using HadCM3LC with climate projections from five Earth system models under SSP2-4.5, SSP5-8.5 and G6sulfur. We first characterise the environmental conditions associated with equilibrium forest persistence before evaluating how future climate states evolve relative to these conditions (see Methods, Section~\ref{sec:sec2}).

\begin{figure}[H]
\noindent\includeoriginal[width=1\linewidth]{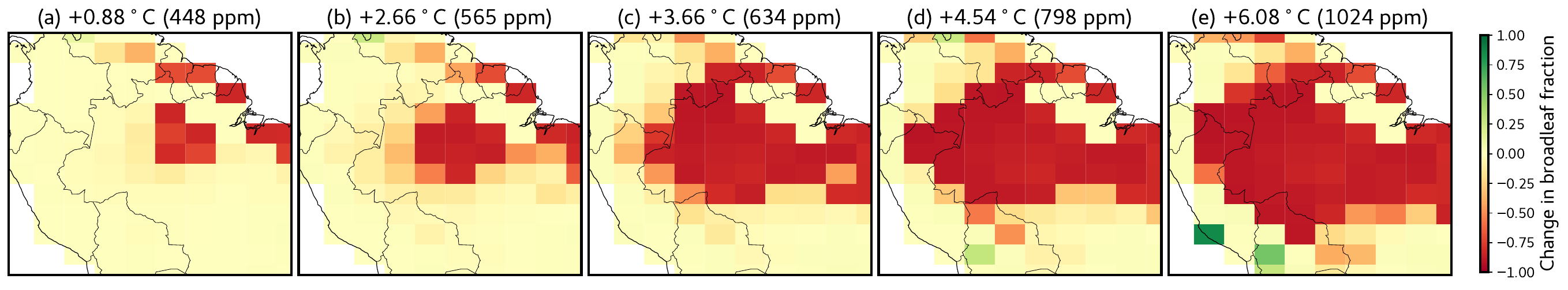}

\caption{Changes in equilibrium broadleaf forest fraction over the Amazon under increasing global warming levels and atmospheric CO$_2$ concentrations in HadCM3LC. Each panel shows the change relative to the reference equilibrium state (CO$_2$ = 389 ppm).The subsequent equilibrium states correspond to
(a) +0.88$^\circ$C (448 ppm), (b) +2.66$^\circ$C (565 ppm), (c) +3.66$^\circ$C (634 ppm), (d) +4.54$^\circ$C (798 ppm), and (e) +6.08$^\circ$C (1024 ppm).
Warming levels represent global-mean near-surface air temperature changes relative to the 389 ppm reference simulation. Negative values indicate a reduction in broadleaf forest fraction, while positive values indicate an increase.}
\label{fig:Hadcm3_dieback}
\end{figure}

Equilibrium simulations show progressively widespread losses of Amazon broadleaf forest cover with increasing global warming (Fig.~\ref{fig:Hadcm3_dieback}). At a global warming level of +3.66~$^\circ$C, corresponding to an atmospheric CO$_2$ concentration of 634~ppm, substantial reductions in broadleaf forest fraction emerge across the Amazon. Forest losses become more extensive at warming levels of +4.54~$^\circ$C and +6.08~$^\circ$C. These changes reflect the combined influence of warming, associated hydroclimatic changes and increasing atmospheric CO$_2$ concentrations, which can also affect vegetation through physiological responses.

To characterise the environmental conditions associated with these forest changes, we examine equilibrium vegetation states in a reduced-dimensional climate--CO$_2$ phase space defined by temperature, dry-season length and atmospheric CO$_2$ concentration. Forest persistence is characterised using the dry-season resilience (DSR) diagnostic, which combines these environmental variables into a common forest sustainability threshold (see Methods). Values of DSR below zero indicate conditions associated with higher equilibrium broadleaf forest cover, whereas positive values indicate conditions associated with lower forest cover.

\begin{figure}[H]
\centering
\includeoriginal[width=1\textwidth]{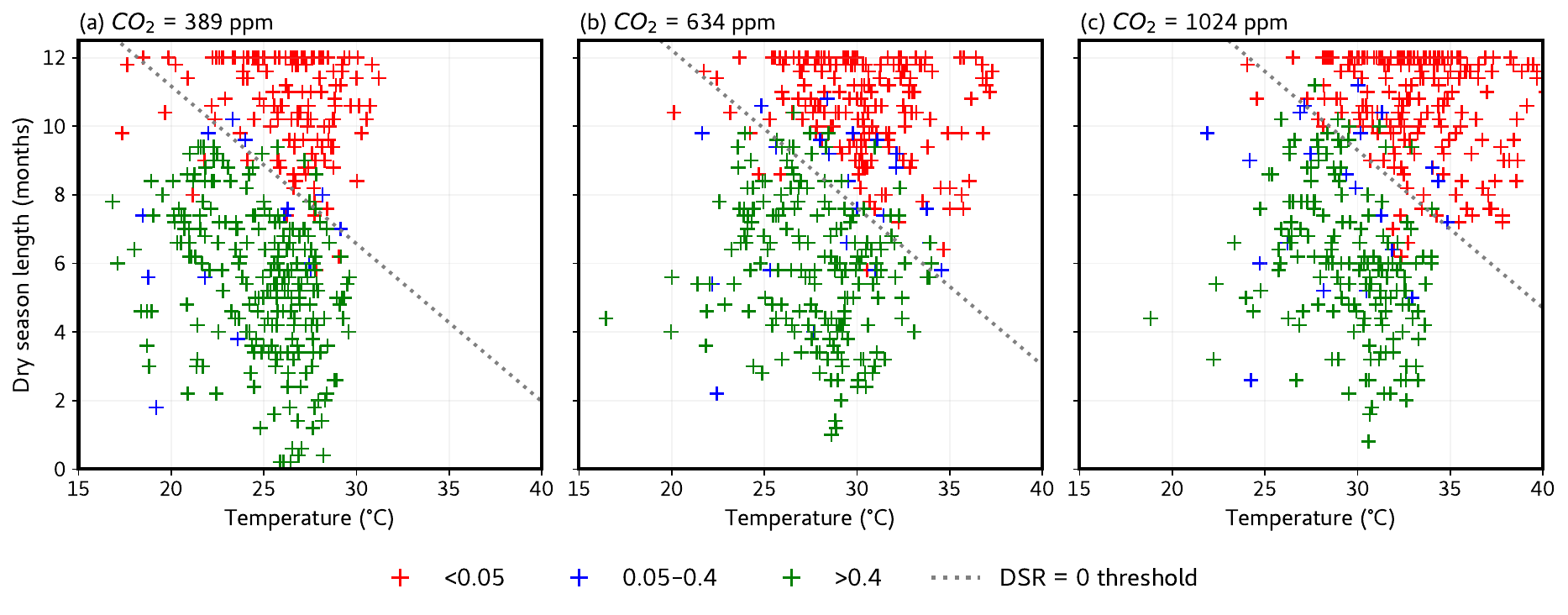}
\caption{
Simulated bioclimatic zones of Amazon equilibrium broadleaf forest fraction under different atmospheric CO$_2$ concentrations. Each panel shows broadleaf forest fraction, indicated by symbol colour, as a function of dry-season length and surface air temperature. Each cross represents one Amazon grid cell from the corresponding equilibrium simulation. Red crosses indicate broadleaf forest fraction $<0.05$, blue crosses indicate intermediate broadleaf fraction between $0.05$ and $0.4$, and green crosses indicate broadleaf fraction $>0.4$. The grey dotted line shows the $\mathrm{DSR}=0$ threshold for (a) CO$_2$ = 389 ppm, (b) CO$_2$ = 634 ppm, and (c) CO$_2$ = 1024 ppm.
}
\label{fig:amazon_bioclimatic_zones}
\end{figure}

The distribution of equilibrium forest states in temperature--dry-season-length space reveals a relationship between forest cover and environmental conditions (Fig.~\ref{fig:amazon_bioclimatic_zones}). Grid cells characterised by higher temperatures and longer dry seasons generally exhibit lower broadleaf forest fractions. However, the position of the DSR = 0 threshold also shifts with increasing atmospheric CO$_2$, towards conditions with greater temperature and dry-season stress. Thus, increasing CO$_2$ modifies the environmental conditions associated with forest persistence, even as the simulated climate shifts towards warmer and seasonally drier states.

Expressing forest persistence in this common climate--CO$_2$ phase space enables projected climate states from different Earth system models to be evaluated against the same environmental threshold, independently of the geographical patterns of forest change simulated by HadCM3LC. We use this framework to assess how projected warming, dry-season changes and elevated atmospheric CO$_2$ influence exposure to conditions associated with lower forest cover.

Good et al. (2011)\cite{good2011quantifying} evaluated the HadCM3 threshold for present day CO$_2$ levels against observations. The model behaved well, although had a tendency towards too binary a response, with much fewer gridpoints being of partial forest cover. But the distinction between forest and non-forest matched observed patterns of the worlds tropical forest well. This gives us confidence in the baseline threshold under present day conditions, but it should be noted that the CO$_2$-dependence of the threshold is not observationally constrained and thus depends only on the model's CO$_2$ response of vegetation. HadCM3LC does, though, have a CO$_2$ fertilisation effect in the middle of the range of coupled climate-carbon cycle models (termed "beta", \cite{jones2020quantifying}).

\subsection{Amazon climate changes across forcing pathways}

\begin{figure}[H]
\centering
\includeoriginal[width=1\linewidth]{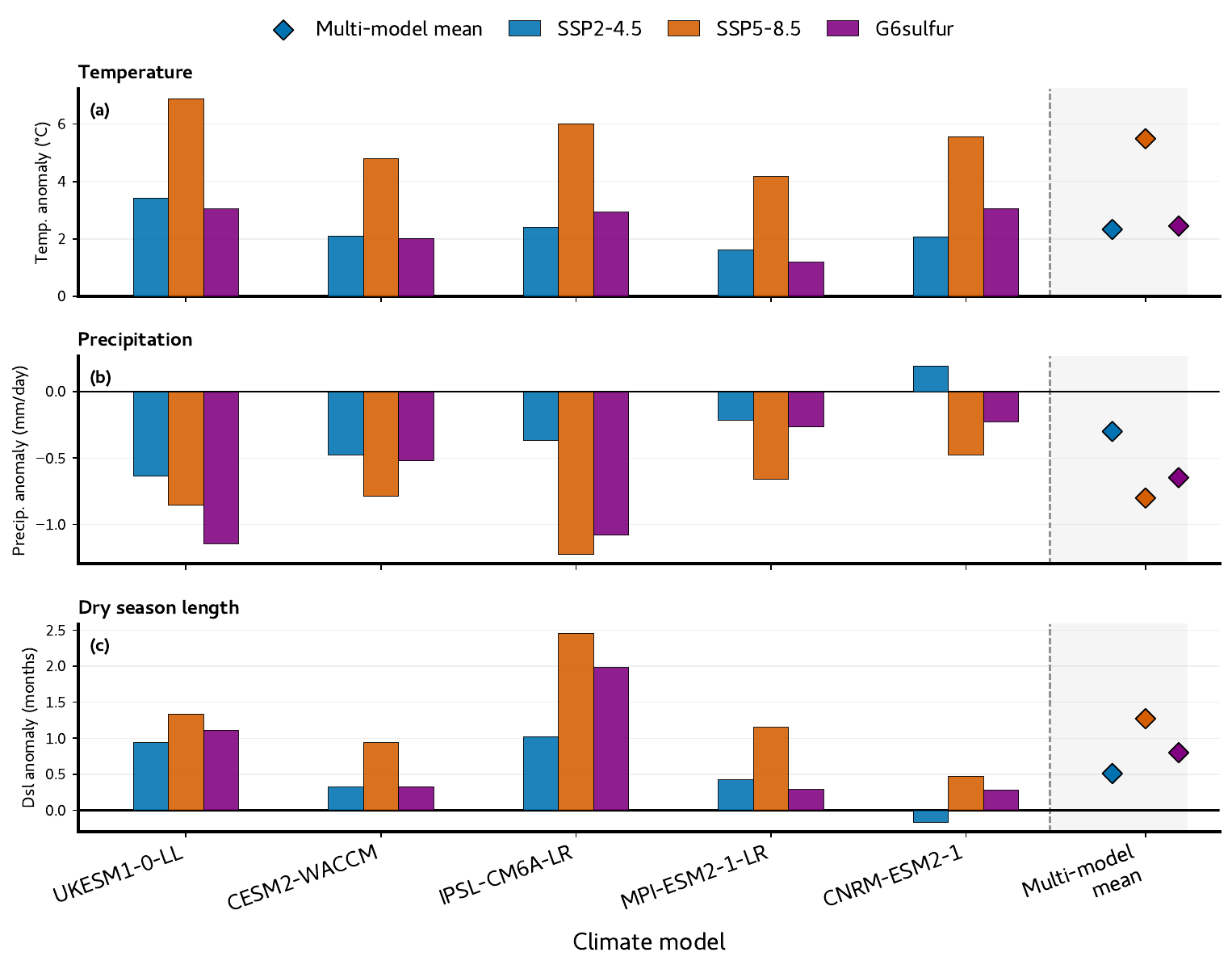}
\caption{Projected late century changes in Amazon climatic conditions across five Earth system models under SSP2-4.5, SSP5-8.5, and G6sulfur. Anomalies are calculated as the difference between the end of century period (2091 - 2100) and the present day period (2015 - 2025) for (a) near surface air temperature, (b) precipitation, and (c) dry-season length. Bars show the Amazon domain mean anomaly for each individual model and scenario, while diamonds show the corresponding multi-model means. Positive temperature and dry-season length anomalies indicate warming and dry-season lengthening, respectively, while negative precipitation anomalies indicate drying.}
\label{fig:multimodel_mean_anomalies}
\end{figure}

We first examine how the climatic conditions relevant to Amazon forest persistence change under mitigation, high-emission and SAI pathways. Figure~\ref{fig:multimodel_mean_anomalies} compares late-century changes in Amazon temperature, precipitation and dry-season length across five Earth system models under SSP2-4.5, SSP5-8.5 and G6sulfur.

All five models project substantial Amazon warming under SSP5-8.5, with a multi-model mean increase of $5.49~^\circ$C (Fig.~\ref{fig:multimodel_mean_anomalies}a). G6sulfur reduces this warming in every model, producing a multi-model mean increase of $2.45~^\circ$C, close to the $2.33~^\circ$C under SSP2-4.5. The magnitude of warming nevertheless varies among models, with UKESM1-0-LL showing the largest increase under SSP5-8.5 and MPI-ESM1-2-LR showing the smallest warming across all three scenarios.

The precipitation response differs substantially from that of temperature
(Fig.~\ref{fig:multimodel_mean_anomalies}b). All five models project drying under SSP5-8.5, with a multi-model mean precipitation anomaly of $-0.80~\mathrm{mm~day^{-1}}$. Under G6sulfur, drying is reduced relative to SSP5-8.5 in CESM2-WACCM, MPI-ESM1-2-LR and CNRM-ESM2-1, remains similar in IPSL-CM6A-LR, and becomes stronger in UKESM1-0-LL. Consequently, the multi-model mean precipitation anomaly under G6sulfur remains $-0.65~\mathrm{mm~day^{-1}}$, compared with $-0.30~\mathrm{mm~day^{-1}}$ under SSP2-4.5. This contrast is consistent with the different responses of the hydrological cycle to greenhouse-gas and solar radiative forcing \citep{kravitz2013energetic,haywood2013asymmetric}.

Dry-season length similarly shows only a partial offset of the SSP5-8.5 response (Fig.~\ref{fig:multimodel_mean_anomalies}c).
Under SSP5-8.5, the dry season lengthens in all five models, with a multi-model mean increase of 1.27 months. G6sulfur reduces this increase
in every model, giving a multi-model mean of 0.80 months, compared with 0.51 months under SSP2-4.5. However, substantial inter-model differences remain. CESM2-WACCM and MPI-ESM1-2-LR show G6sulfur responses close to, or smaller than, those under SSP2-4.5, whereas IPSL-CM6A-LR retains an increase of almost two months. UKESM1-0-LL also shows greater dry-season lengthening under G6sulfur than under SSP2-4.5.

Changes in mean precipitation do not necessarily translate into proportional changes in dry-season length. Because dry-season length is defined as the number of months with precipitation below 100~mm~month$^{-1}$, reductions in rainfall do not always result in additional months crossing this threshold. This distinction helps explain why G6sulfur offsets dry-season lengthening more strongly than mean precipitation losses.

To quantify these contrasting responses, we calculate the percentage of the difference between SSP5-8.5 and SSP2-4.5 anomalies offset under G6sulfur (Methods; Eq.~\ref{eq:climate_recovery}). In the multi-model mean, G6sulfur closes approximately 96\% of this difference in temperature, but only 30\% in precipitation and 62\% in dry-season length. Thus, although SAI brings Amazon warming close to SSP2-4.5 levels, substantial hydroclimatic changes persist. These differences raise the question of whether reducing warming alone is sufficient to limit exposure to climatic conditions associated with long-term forest loss.

\subsection{Amazon climate states relative to forest-sustainability thresholds}

\begin{figure}[H]
\noindent\includeoriginal[width=.8\textwidth]{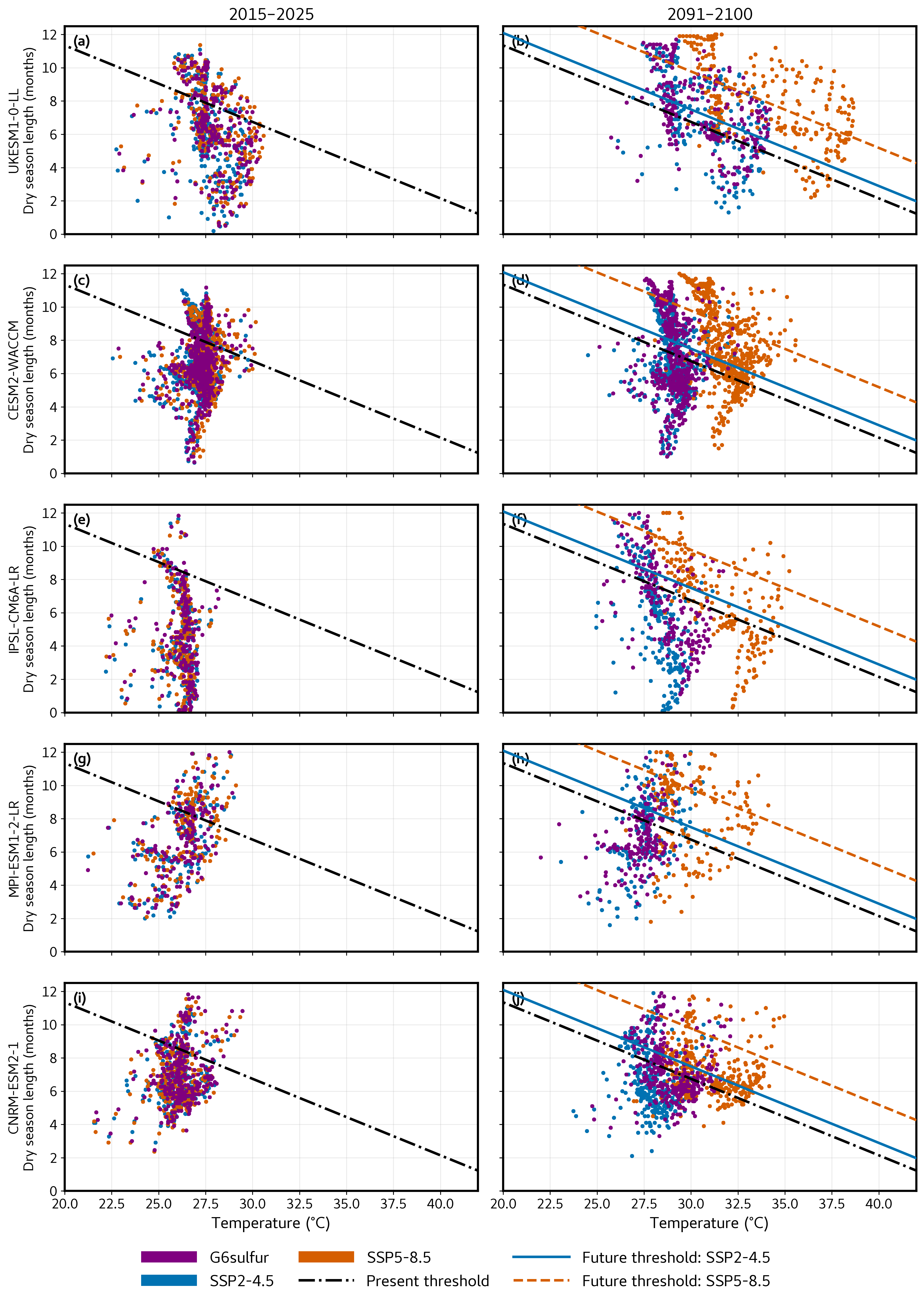}
\caption{Projected Amazon climate states relative to CO$_2$-dependent forest-sustainability thresholds across five Earth system models. Each row represents an individual model, with near-present (2015--2025) climate conditions shown in the left column and late-century (2091--2100) conditions in the right column. Each point represents an Amazon land grid cell positioned according to near-surface air temperature and dry-season length. The distributions show climate states under SSP2-4.5,
SSP5-8.5 and G6sulfur. The forest-sustainability threshold (DSR = 0) separates conditions associated with higher and lower equilibrium broadleaf forest cover. The near-present threshold corresponds to 429~ppm CO$_2$, while late-century scenario-specific thresholds correspond to 603~ppm under SSP2-4.5 and 1135~ppm under SSP5-8.5 and G6sulfur.}
\label{fig:multimodel_scatter}
\end{figure}

To assess how the contrasting temperature and hydroclimatic responses influence conditions associated with forest persistence, we examine the distribution of Amazon grid cells in temperature--dry-season-length phase space relative to the forest-sustainability threshold (Fig.~\ref{fig:multimodel_scatter}). During the near-present period (2015--2025), the three scenarios exhibit similar climate conditions and atmospheric CO$_2$ concentrations (429~ppm), resulting in closely overlapping distributions and a common present-day threshold.

By late century (2091--2100), projected warming and dry-season lengthening shift Amazon climate states towards conditions associated with lower forest cover. The displacement is generally greater under SSP5-8.5 than under SSP2-4.5, reflecting their contrasting temperature and hydroclimatic responses (Fig.~\ref{fig:multimodel_mean_anomalies}). However, the forest-sustainability threshold also shifts with atmospheric CO$_2$ concentration. By late century, the threshold corresponds to 603~ppm under SSP2-4.5 and 1135~ppm under SSP5-8.5, with higher CO$_2$ shifting the boundary towards conditions of greater climatic stress.

G6sulfur produces a distinct combination of climatic and physiological conditions. Although SAI reduces Amazon warming substantially relative to SSP5-8.5, it retains the same high atmospheric CO$_2$ concentration (1135~ppm), while residual drying and dry-season lengthening persist. Consequently, Amazon climate states under G6sulfur are generally displaced towards less stressful temperature--dry-season-length conditions than under SSP5-8.5, while being evaluated against the same future CO$_2$-dependent forest-sustainability threshold.

The resulting threshold-exceedance diagnostic therefore reflects two
counteracting influences: warming and dry-season lengthening shift climate states towards conditions associated with lower forest cover, whereas elevated CO$_2$ shifts the threshold towards greater climatic stress in the underlying vegetation framework. Comparing future climate states against both the present-day and scenario-specific future thresholds allows the contribution of the CO$_2$-dependent threshold shift to be distinguished from the effect of changes in climate alone. The corresponding differences in threshold exceedance are quantified in the following section.

\subsection{CO$_2$ dependence of diagnosed threshold exceedance}

\begin{figure}[H]
\noindent\includeoriginal[width=1\textwidth]{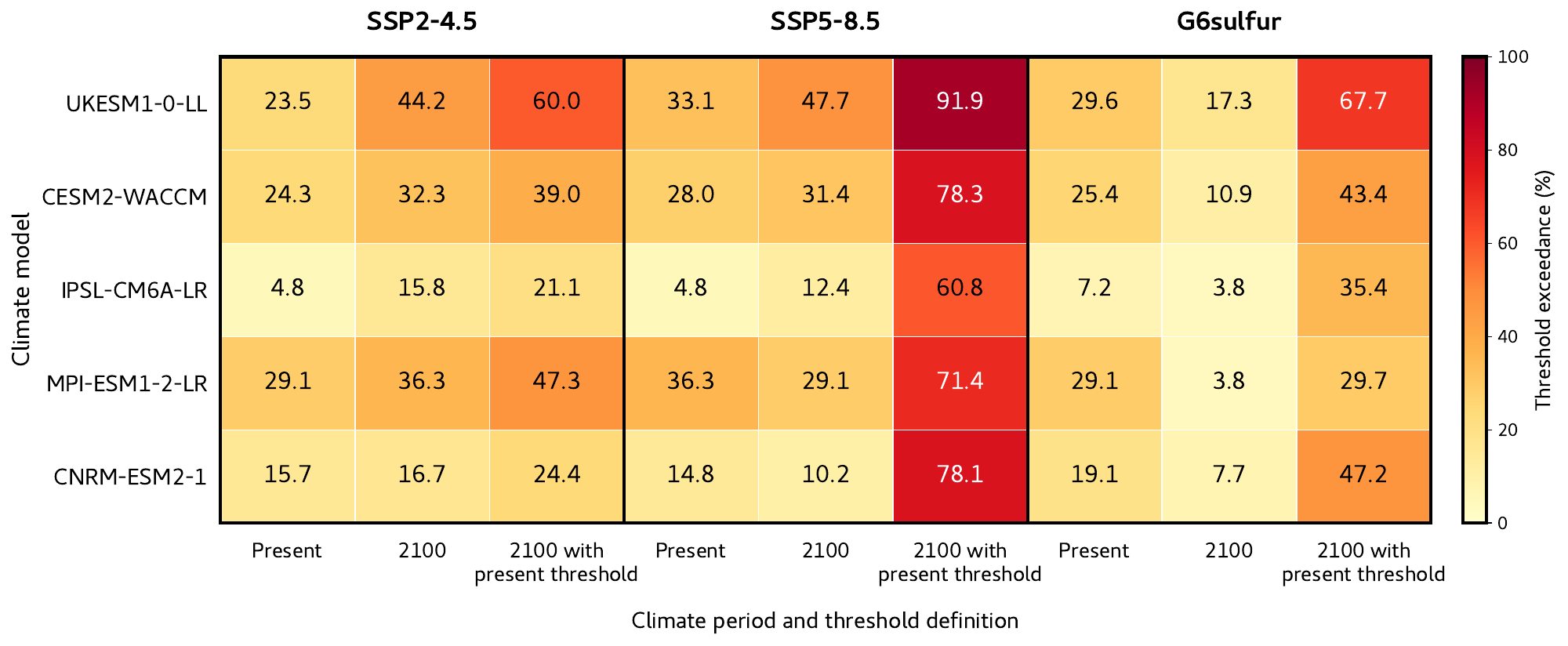}
\caption{Percentage of Amazon grid cells crossing the CO$_2$ dependent dieback threshold across models, scenarios, time periods, and threshold definitions. For each scenario, the three columns represent:
(i) near-present climate (2015--2025) evaluated against the present-day CO$_2$ threshold;
(ii) late-century climate (2091--2100) evaluated against the corresponding future
scenario-specific CO$_2$ threshold; and
(iii) late-century climate (2091--2100)
evaluated against the fixed present-day CO$_2$ threshold.}
\label{fig:multimodel_mean_ppercentage}
\end{figure}

Figure~\ref{fig:multimodel_mean_ppercentage} quantifies the fraction of Amazon grid cells exceeding the forest-sustainability threshold across models and forcing pathways. When late-century climate states are evaluated against the present-day CO$_2$ threshold, exceedance is greatest under SSP5-8.5 (76.1\%), followed by G6sulfur (44.7\%) and SSP2-4.5 (38.4\%) in the multi-model mean. Thus, SAI substantially reduces exposure to threshold-exceeding conditions relative to SSP5-8.5, although exceedance remains greater than under SSP2-4.5 when the CO$_2$ threshold is held constant.

Accounting for the scenario-specific future CO$_2$ threshold substantially changes this assessment. Mean exceedance decreases to 26.2\% under SSP5-8.5, 8.7\% under G6sulfur and 29.1\% under SSP2-4.5. Notably, G6sulfur now exhibits the lowest threshold exceedance among the three scenarios, reflecting the combination of reduced warming and the high-CO$_2$ forest-sustainability threshold.

The contrast between the two threshold definitions demonstrates that inferred forest vulnerability depends strongly on the CO$_2$-dependent vegetation response represented in the diagnostic. Under G6sulfur, exceedance decreases from 44.7\% to 8.7\% when the future rather than present-day threshold is applied, despite identical underlying climate conditions.

\section{Discussion}

Our results demonstrate that the potential of stratospheric aerosol injection (SAI) to reduce Amazon forest vulnerability cannot be inferred from temperature reduction alone. Although G6sulfur closes approximately 96\% of the difference in Amazon warming between SSP5-8.5 and SSP2-4.5, the corresponding recovery is substantially smaller for precipitation (30\%) and dry-season length (62\%). These contrasting responses are particularly important for Amazon forest persistence, which depends strongly on dry-season water availability \citep{good2011quantifying,levine2016ecosystem}. Previous projections of Amazon dry-season intensification further highlight the importance of hydroclimatic changes for forest vulnerability \citep{boisier2015projected}.

The forest-sustainability threshold analysis provides a complementary perspective on the potential benefits of SAI. Using a fixed present-day CO$_2$ threshold, G6sulfur reduces the multi-model mean threshold exceedance from 76.1\% under SSP5-8.5 to 44.7\%, although this remains above the 38.4\% under SSP2-4.5. However, applying scenario-specific future CO$_2$ thresholds reduces exceedance under G6sulfur to 8.7\%, compared with 29.1\% under SSP2-4.5, reversing their relative ranking. This illustrates the strong dependence of inferred forest protection on the CO$_2$-dependent vegetation response. Previous studies have suggested that SAI could increase Amazon carbon storage through reduced warming and elevated atmospheric CO$_2$ \citep{parry2026stratospheric}. Our analysis extends this perspective by evaluating the environmental conditions associated with long-term forest persistence, rather than carbon storage alone.

Sensitivity experiments further demonstrate that the magnitude of threshold exceedance depends on the assumed strength of the CO$_2$-dependent vegetation response (Supplementary Fig.~S3). Although the relative ranking of SSP2-4.5 and G6sulfur changes with this assumption, G6sulfur consistently exhibits lower threshold exceedance than SSP5-8.5 across all five models and the tested CO$_2$-response strengths (Supplementary Fig.~S4). This suggests that the relative reduction in threshold exposure under SAI persists across these sensitivity experiments, even though the magnitude of the inferred benefit remains dependent on the CO$_2$ response.

Expressing forest-sustainability conditions in a reduced-dimensional climate-CO$_2$ phase space provides a common diagnostic for evaluating projected climate states across Earth system models, without directly transferring geographical patterns of forest loss from HadCM3LC. The CO$_2$-dependent threshold response remains consistent across the HadCM3LC equilibrium experiments, including at elevated CO$_2$ concentrations, supporting its application to the future scenarios considered here. Furthermore, the present-day threshold has previously been evaluated against observed tropical forest distributions \citep{good2011quantifying}. Nevertheless, the magnitude of the CO$_2$ dependent vegetation response remains observationally unconstrained, and biases in simulated climate may affect threshold classifications across Earth system models.

Although the $ DSR = 0$ threshold is derived from equilibrium simulations with HadCM3LC, generating equivalent multi-century equilibrium ensembles across modern CMIP6 Earth system models is computationally expensive. Among the five models evaluated here, only UKESM1-0-LL and MPI-ESM1-2-LR includes interactive dynamic vegetation. UKESM1-0-LL is coupling an updated version of the same TRIFFID framework used in HadCM3LC within the JULES-ES land-surface scheme. Whereas HadCM3LC employs an earlier five-plant-functional-type configuration without nutrient constraints, UKESM1-0-LL includes expanded plant functional types, updated trait parameterisations, and an interactive terrestrial nitrogen cycle that dampens CO$_2$-driven fertilisation. Because nutrient limitations and structural differences across dynamic vegetation models can attenuate the physiological resilience gained from elevated CO$_2$, the HadCM3LC-derived threshold shift may represent an upper-end response. Our $\alpha$-scaling sensitivity analysis explicitly accounts for this structural uncertainty by evaluating how Amazon threshold exposure responds across weaker-to-stronger CO$_2$ fertilisation regimes, while a comprehensive assessment of inter-model threshold uncertainty remains an important priority for future work.

Forest persistence is also influenced by processes not explicitly represented in this diagnostic, including soil-water availability, fire, deforestation and vegetation--atmosphere feedbacks \citep{levine2016ecosystem,brando2025tipping}. Observational evidence of declining Amazon resilience, particularly in drier regions, highlights the importance of environmental stress beyond changes in forest cover \citep{boulton2022pronounced}. Recent evidence that deforestation-induced drying can lower the climatic threshold for Amazon forest transitions further emphasises the importance of moisture-recycling feedbacks \citep{wunderling2026deforestation}. Consequently, threshold exceedance should be interpreted as exposure to environmental conditions associated with lower equilibrium forest cover, rather than a direct prediction of forest dieback.

Finally, G6sulfur represents one hypothetical SAI pathway under high atmospheric CO$_2$, and its outcomes should not be generalised to all geoengineering strategies. Long-term forest responses may also persist beyond periods of elevated warming, as demonstrated in studies of committed ecosystem change and temperature overshoot \citep{jones2009committed,munday2025risks}. Our findings suggest that SAI can reduce exposure to conditions associated with Amazon forest loss, but the extent of this benefit depends on residual hydroclimatic changes and CO$_2$-dependent vegetation responses. Constraining these responses is essential for assessing the potential and limitations of solar geoengineering for tropical forest protection.

\section{Methods}
\label{sec:sec2}

\subsection{HadCM3LC equilibrium simulations}

Most CMIP ESMs do not include vegetation dynamics, and those which do are seldome run to very long-term equilibrium states. Therefore in order to derive a threshold we use a model with this capability. We use simulations from HadCM3LC to define the climatic conditions associated with long-term Amazon forest persistence. HadCM3LC is a coupled climate-carbon cycle general circulation model that participated in C$^4$MIP \citep{friedlingstein2006climate}. The model couples the MOSES2 land-surface scheme \citep{essery2003explicit} with the dynamic vegetation model TRIFFID \citep{cox2001description}, which simulates changes in the fractional coverage of broadleaf trees and other plant functional types in response to climate and atmospheric CO$_2$.

While the modern CMIP6 model UKESM1-0-LL couples an updated version of TRIFFID within the JULES-ES land-surface scheme, incorporating expanded plant functional types and an interactive terrestrial nitrogen cycle, running comparable multi-century equilibrium simulations with higher-complexity Earth system models remains computationally prohibitive.

HadCM3LC is particularly suitable for defining an Amazon forest threshold for three reasons. First, it includes dynamic vegetation, allowing broadleaf forest cover to respond interactively to changing environmental conditions. Second, equilibrium vegetation experiments are available, allowing the long term forest state associated with a given climate and atmospheric CO$_2$ concentration to be diagnosed rather than relying only on the transient vegetation response. This is important because vegetation can adjust more slowly than the atmospheric climate, such that transient simulations may represent only part of the committed vegetation change \citep{jones2009committed, good2013comparing}. Third, HadCM3LC is fully coupled, allowing land--atmosphere feedbacks associated with forest loss to influence the simulated climate and vegetation response.

We use HadCM3LC equilibrium simulations initialized from different stages of a transient experiment in which atmospheric CO$_2$ increases at 1\% yr$^{-1}$ \citep{jones2009committed, good2013comparing}. The equilibrium states correspond to atmospheric CO$_2$ concentrations of 389, 448, 565, 634, 798, and 1024 ppm. For each state, we use equilibrium broadleaf forest fraction together with annual-mean near-surface air temperature and dry-season length (DSL), defined as the number of months per year with precipitation below 100 mm month$^{-1}$.

\subsection{Equilibrium forest dieback threshold}
\label{subsec:threshold}

To identify climatic conditions associated with equilibrium Amazon forest loss, we adopt the dry-season resilience (DSR) framework previously developed for HadCM3LC \citep{good2011quantifying}. The diagnostic combines annual-mean near-surface air temperature ($T$), dry-season length (DSL), and atmospheric CO$_2$ concentration:

\begin{equation}
\mathrm{DSR} =
\mathrm{DSL} + 0.46T - 0.0043\mathrm{CO}_2 - 18.7 .
\end{equation}

The $\mathrm{DSR}=0$ contour represents the threshold separating climate-CO$_2$ conditions associated with high and low equilibrium broadleaf forest cover. Negative values lie on the forest-sustaining side of the threshold, whereas positive values indicate conditions associated with low equilibrium broadleaf forest fraction.

\subsection{Application of the HadCM3LC threshold to multi-model projections}

We apply this forest-sustainability threshold as a diagnostic framework to assess projected Amazon climate states across five Earth system models: UKESM1-0-LL, CESM2-WACCM, CNRM-ESM2-1, IPSL-CM6A-LR, and MPI-ESM1-2-LR.

For each model, monthly near-surface air temperature and precipitation are analysed under three experiments: SSP2-4.5 and SSP5-8.5 from CMIP6 ScenarioMIP, and G6sulfur from GeoMIP. G6sulfur is a stratospheric aerosol intervention experiment based on the SSP5-8.5 forcing pathway.

Rather than using each model's simulated vegetation response, Amazon grid cells from each experiment are projected onto the same temperature and dry-season-length climate space and evaluated relative to the threshold. This provides a common diagnostic for comparing how projected changes in temperature, dry-season length, and atmospheric CO$_2$ move Amazon climate states towards or away from conditions associated with sustainable forest cover.

The Amazon domain is defined over $20^{\circ}$S - $10^{\circ}$N and $80^{\circ}$W - $45^{\circ}$W. For every Amazon grid cell, annual mean temperature and dry season length are calculated. These grid cell values are then averaged over the selected analysis periods to represent the near-present (2015 - 2025) and late-century (2091 - 2100) climate states. Each Amazon grid cell is then positioned in temperature-dry season length phase space and compared against the forest resilience threshold for three different scenarios.

To quantify threshold exceedance, we calculate the fraction of Amazon grid cells crossing the threshold for each model, scenario, and analysis period. We then examine late-century changes in the key climatic drivers by comparing the near-present period (2015 - 2025) with the late-century period (2091 - 2100) for temperature, precipitation, and dry-season length. 
As the G6sulfur experiment begins in 2020 and branches from SSP5-8.5, its near-present reference climate (2015--2025) is taken from the corresponding SSP5-8.5 simulation. Late-century G6sulfur anomalies are therefore calculated relative to this reference,
ensuring a consistent comparison period across the scenarios.

To quantify the extent to which stratospheric aerosol injection offsets regional climate changes, we define a recovery metric for each climate variable $X$:

\begin{equation}
R_X = 100 \times
\frac{
\Delta X_{\mathrm{SSP5-8.5}} -
\Delta X_{\mathrm{G6sulfur}}
}{
\Delta X_{\mathrm{SSP5-8.5}} -
\Delta X_{\mathrm{SSP2-4.5}}
}.
\label{eq:climate_recovery}
\end{equation}

Here, $\Delta X$ represents the late-century (2091--2100) anomaly relative to the near-present (2015--2025) period, and $R_X$ represents the percentage of the difference between SSP5-8.5 and SSP2-4.5 anomalies offset under G6sulfur.
A value of 100\% indicates that the G6sulfur anomaly equals the SSP2-4.5 anomaly, whereas 0\% indicates no offset relative to SSP5-8.5. The metric is calculated separately for temperature, precipitation and dry-season length using their
multi-model mean anomalies.

\section*{Data availability}
The HadCM3LC equilibrium simulations used to define the forest-sustainability threshold are from previously published work \citep{jones2009committed,good2011quantifying}. The CMIP6 ScenarioMIP and GeoMIP model outputs analysed in this study are publicly available through the Earth System Grid Federation (ESGF). Details of the models and experiments used are provided in the manuscript and accompanying code repository.

\section*{Code availability}
The Python code used for the CMIP6/GeoMIP data processing, calculation of temperature, precipitation and dry-season length anomalies, application of the CO$_2$ dependent forest-dieback threshold, threshold crossing analysis, and generation of the figures presented in this study is publicly archived on Zenodo at \url{https://doi.org/10.5281/zenodo.23020320} last access: 28/09/2026.

\section*{Acknowledgements}

This work was supported by UK Research and Innovation (UKRI) through Natural Environment Research Council (NERC) funding under the project `Advancing holistic risk assessment for measures to address climate change' (Project Reference UKRI1025). The authors acknowledge the World Climate Research Programme for coordinating and promoting CMIP6 and thank the participating climate modelling groups for producing and making their model output available. We also acknowledge the Earth System Grid Federation (ESGF) for providing access to the CMIP6 and GeoMIP datasets used in this study.

\section*{Author contributions}

AKX contributed to the conceptualization of the study, development and implementation of the methodology, data analysis, visualization, and preparation of the manuscript. CJ contributed to the conceptualization of the study, supervision, method development, and critical review and improvement of the manuscript.

\section*{Competing interests}

The authors declare that they have no conflict of interest.

\bibliographystyle{naturemag}
\bibliography{agusample}
\end{document}


\maketitle
\vspace{-1.4em}

\section*{Overview}
This document provides additional model-resolved diagnostics and a sensitivity assessment of the CO$_2$-dependent forest-sustainability threshold used in the main manuscript. Threshold exceedance diagnoses environmental conditions associated with reduced equilibrium broadleaf forest cover; it is \emph{not} a direct simulation of future forest loss.

\section*{Model-specific threshold exceedance}
The model-resolved comparison in Supplementary Fig.~\ref{fig:s1} complements the scenario-mean threshold-exceedance results in the main manuscript. It shows how the near-present and late-century exceedance fractions differ among the five Earth system models when each scenario is evaluated using its corresponding CO$_2$-dependent threshold.

\begin{figure}[H]
\centering
\includegraphics[width=\linewidth]{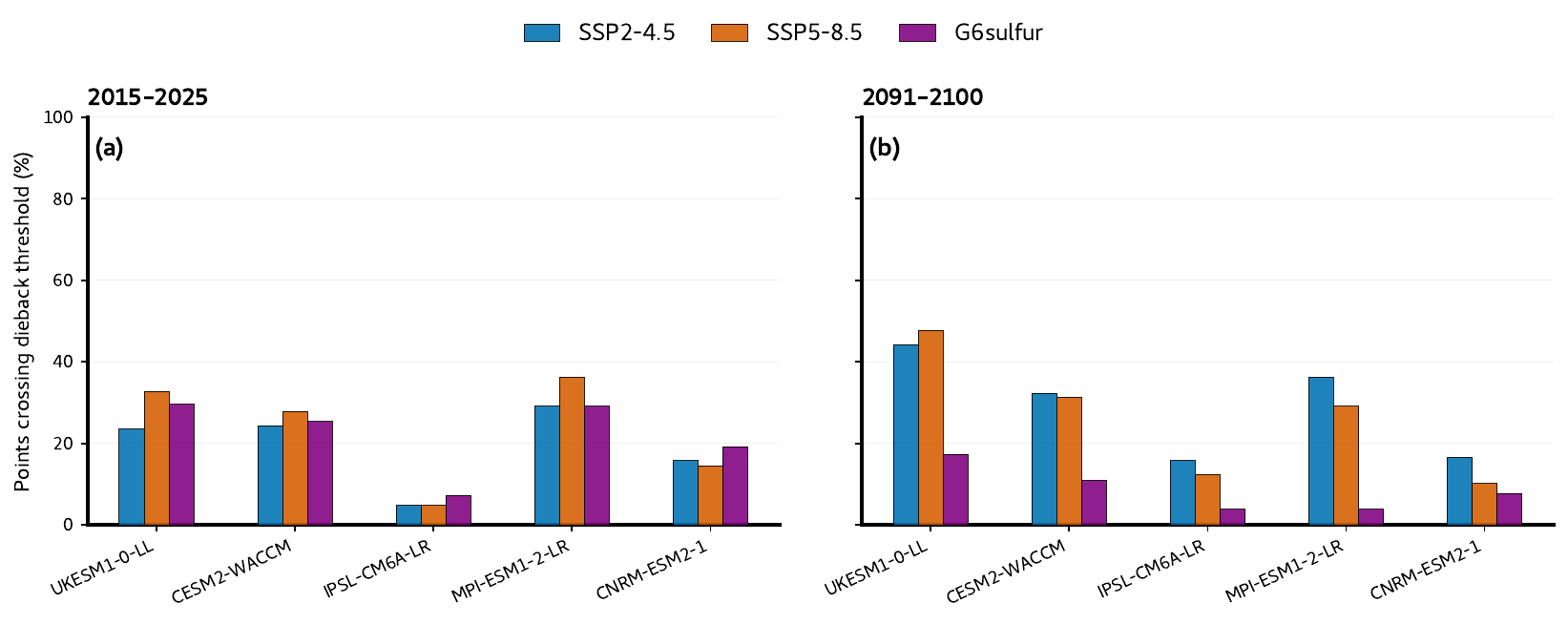}
\caption{\textbf{Threshold exceedance in individual Earth system models.} Percentage of analysed Amazon grid cells exceeding the CO$_2$-dependent forest-sustainability threshold under SSP2-4.5, SSP5-8.5 and G6sulfur. (a) Near-present climate (2015--2025) and (b) late-century climate (2091--2100), each evaluated against the CO$_2$ threshold associated with its scenario and period. The metric identifies threshold-exceeding climate states, not the percentage of actual forest cover lost.}
\label{fig:s1}
\end{figure}

\section*{Model-resolved climate anomalies}
\begin{figure}[H]
\centering
\includegraphics[width=\linewidth]{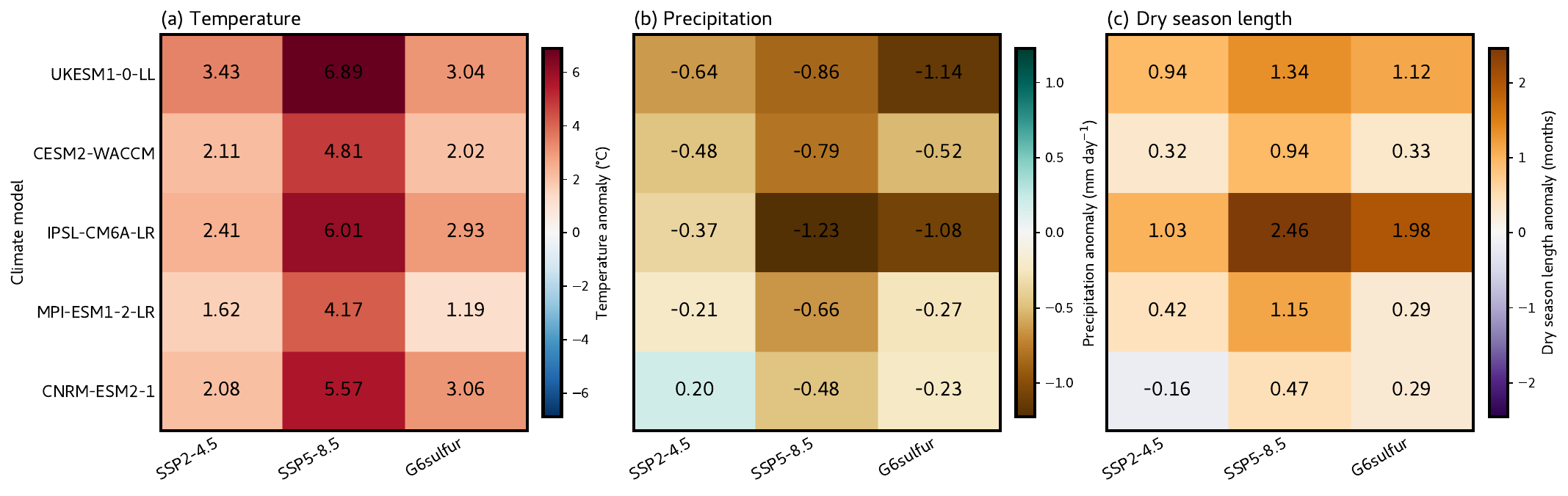}
\caption{\textbf{Model-specific Amazon climate changes.} Late-century (2091--2100) minus near-present (2015--2025) anomalies in (a) near-surface air temperature (\,$^\circ$C), (b) precipitation (mm\,day$^{-1}$) and (c) dry-season length (months). Each cell gives the mean anomaly for the indicated model and scenario. Positive temperature and dry-season-length anomalies indicate warming and lengthening, whereas negative precipitation anomalies indicate drying.}
\label{fig:s2}
\end{figure}

Supplementary Fig.~\ref{fig:s2} reports model-specific anomalies underlying the multi-model climate comparisons. The heatmaps illustrate the contrast between relatively consistent temperature offsets and more heterogeneous precipitation and dry-season responses under G6sulfur.

\clearpage
\section*{Sensitivity to the CO$_2$-dependent threshold response}
To evaluate the dependence of diagnosed exceedance on the CO$_2$ term in the dry-season resilience (DSR) diagnostic, the future CO$_2$ increment was scaled using a dimensionless multiplier $\alpha$ while holding the projected temperature and dry-season length fixed:
\begin{equation}
\mathrm{DSR}_{\alpha} = \mathrm{DSL} + 0.46T
- 0.0043\left[CO_{2,\mathrm{present}} + \alpha\bigl(CO_{2,\mathrm{future}}-CO_{2,\mathrm{present}}\bigr)\right] -18.7.
\label{eq:sensitivity}
\end{equation}
Here $CO_{2,\mathrm{present}}=429.03$~ppm; $CO_{2,\mathrm{future}}=602.78$~ppm for SSP2-4.5 and 1135.21~ppm for SSP5-8.5 and G6sulfur. Values of $\alpha=0,\,0.25,\,0.50,\,0.75,\,1.00$ and $1.25$ were examined. At $\alpha=0$, the future climate is assessed against the present-day CO$_2$ threshold; $\alpha=1$ recovers the original future CO$_2$ shift. These are illustrative sensitivity cases rather than probabilities or an estimated confidence interval.

For each model, scenario and $\alpha$, exceedance was recalculated as the proportion of exported future-climate records with $\mathrm{DSR}_{\alpha}>0$. The thick curves are equal-model means of the five available model fractions; thin curves are individual models and should not be interpreted as uncertainty intervals.

\begin{figure}[H]
\centering
\includegraphics[width=\linewidth]{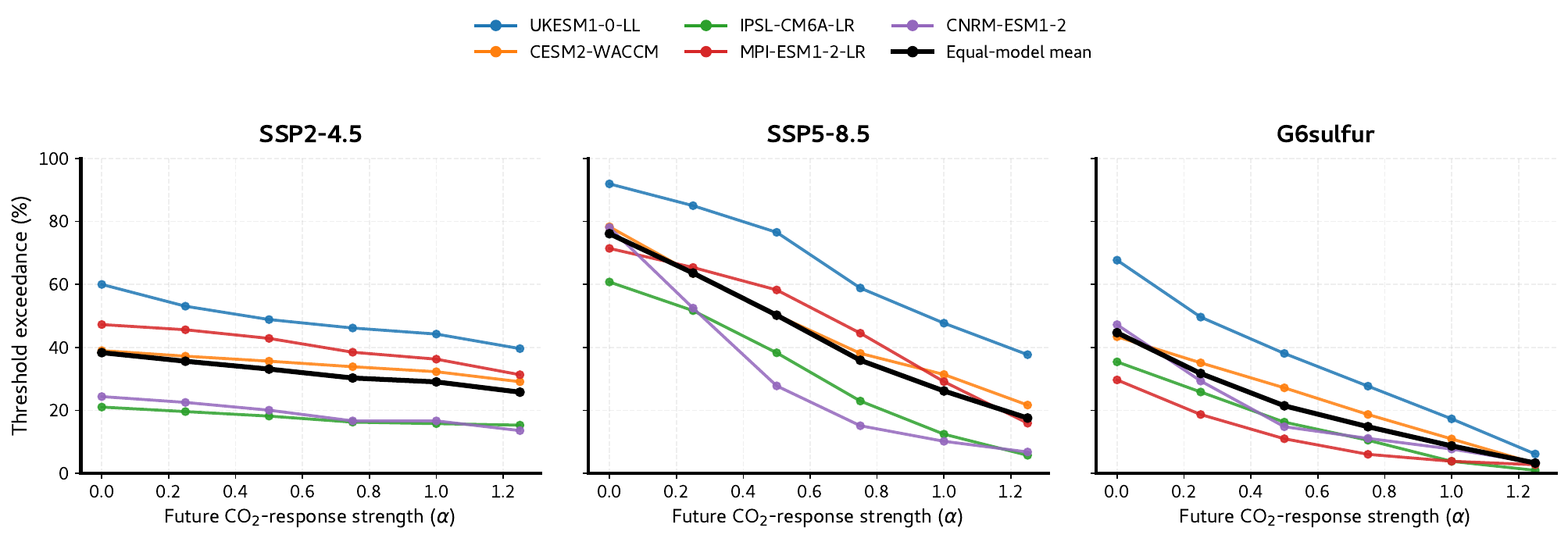}
\caption{
\textbf{Sensitivity of the forest-sustainability diagnostic to the assumed future CO$_2$ response.}
Percentage of the analysed Amazon domain exceeding the
forest-sustainability threshold during the late-century
period (2091--2100) under (left) SSP2-4.5,
(centre) SSP5-8.5 and (right) G6sulfur, as the
strength of the future CO$_2$-dependent threshold
shift is varied using the parameter $\alpha$
(Eq.~\ref{eq:sensitivity}).
Coloured lines represent the five individual Earth system
models, while the thick black lines indicate their
equal-model means.
Temperature and dry-season length remain fixed
at their respective late-century values throughout
the sensitivity experiments.
Threshold exceedance is calculated as the
spatially weighted fraction of the analysed
Amazon domain experiencing conditions beyond
the forest-sustainability threshold.
}
\label{fig:s3}
\end{figure}

\clearpage

\section*{Relative G6sulfur benefit across CO$_2$ assumptions}

SSP5-8.5 and G6sulfur share the same future atmospheric CO$_2$ concentration. Consequently, for each value of $\alpha$, both scenarios are evaluated against the same CO$_2$-dependent forest-sustainability threshold, allowing the effects of their differing
climatic conditions to be compared consistently.

For each Earth system model $m$, the difference in threshold exceedance is calculated as

\begin{equation}
\Delta F_m(\alpha)
=
F_{m,\mathrm{SSP5-8.5}}(\alpha)
-
F_{m,\mathrm{G6sulfur}}(\alpha),
\label{eq:relative_benefit}
\end{equation}

where $F_{m,s}(\alpha)$ represents the spatially weighted percentage of the analysed Amazon domain exceeding the forest-sustainability threshold for model $m$, scenario $s$, and CO$_2$-response strength $\alpha$.

Positive values of $\Delta F_m(\alpha)$ indicate lower threshold exceedance under G6sulfur relative to SSP5-8.5, whereas negative values indicate greater exceedance under G6sulfur. This comparison captures the combined effects of temperature and dry-season-length differences between the scenarios without separately attributing
their individual contributions.

\begin{figure}[H]
\centering
\includegraphics[width=0.88\linewidth]{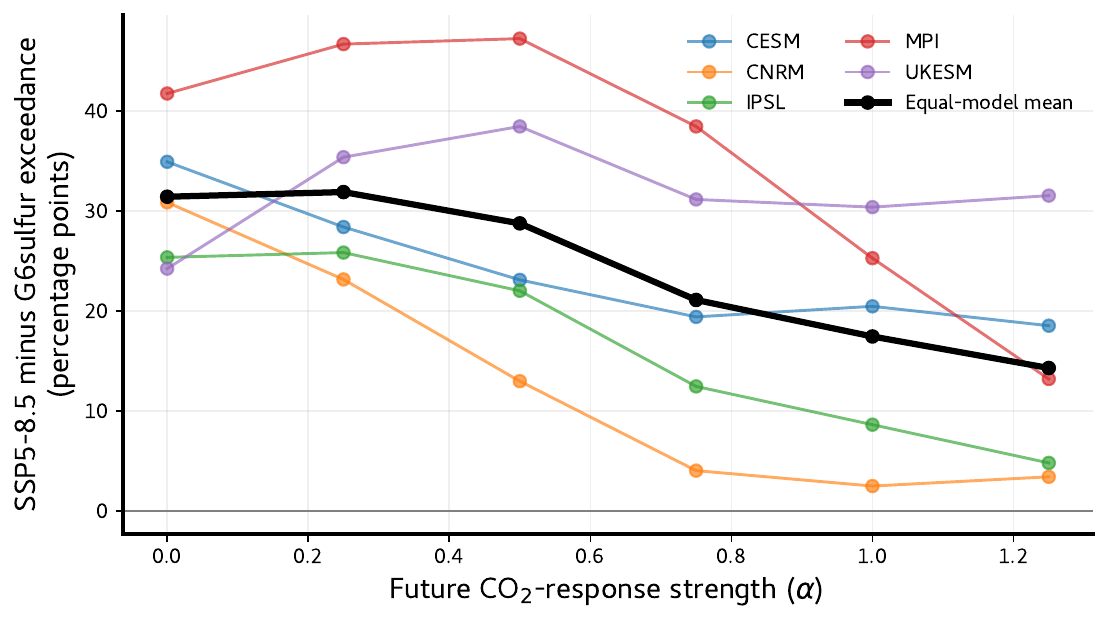}
\caption{
\textbf{Difference in forest-sustainability threshold exceedance between SSP5-8.5 and G6sulfur across assumed CO$_2$ responses.}
For each Earth system model and value of $\alpha$,
the spatially weighted threshold-exceedance
percentage under G6sulfur is subtracted from
that under SSP5-8.5
(Eq.~\ref{eq:relative_benefit}).
Coloured lines represent individual Earth system
models, while the thick black line shows the
equal-model mean.
Both scenarios share the same atmospheric
CO$_2$ concentration and corresponding
forest-sustainability threshold at each
$\alpha$ value.
Positive values indicate a reduction in
threshold exceedance under G6sulfur relative
to SSP5-8.5. These differences represent
changes in exposure to conditions associated
with reduced equilibrium forest cover,
rather than directly simulated forest loss.
}
\label{fig:s4}
\end{figure}